\documentclass[aps,prc,preprint,groupedaddress]{revtex4-1}
\usepackage{graphicx}
\usepackage{amsmath}
\usepackage{amsmath, bm}
\usepackage{ulem}
\usepackage{multirow}

\usepackage[colorlinks,
linkcolor=blue,
anchorcolor=blue,
urlcolor=blue,
citecolor=blue]{hyperref}

\begin{document}

%\preprint{}

\title{Neutron-rich calcium isotopes within realistic Gamow shell model calculations with continuum coupling}


\author{J.G. Li, B.S. Hu, Q. Wu, Y. Gao, S.J. Dai, and F.R. Xu}
%\author{F.R.Xu}
\email[]{frxu@pku.edu.cn}

%\thanks{}
\affiliation{School of Physics,  and   State Key  Laboratory  of  Nuclear  Physics   and  Technology, Peking University, Beijing  100871, China}


\date{\today}

\begin{abstract}

  Based on the realistic nuclear force of the high-precision CD-Bonn potential, we have performed comprehensive calculations for neutron-rich calcium isotopes using the  Gamow shell model (GSM) which includes resonance and continuum. The realistic GSM calculations  produce well  binding energies, one- and two-neutron separation energies, predicting that  $^{57}$Ca is the heaviest bound odd isotope and $^{70}$Ca is the dripline nucleus. Resonant states are predicted, which provides useful information for future experiments on  particle emissions in neutron-rich calcium isotopes. Shell evolutions in the calcium chain around neutron numbers  \textit{N} = 32, 34 and 40 are understood by calculating effective single-particle energies,   the excitation energies of the first  $2^+$ states and two-neutron separation energies. The calculations  support shell closures at $^{52}$Ca (\textit{N} = 32) and $^{54}$Ca (\textit{N} = 34) but show a weakening of shell closure at $^{60}$Ca (\textit{N} = 40). The possible shell closure at $^{70}$Ca (\textit{N} = 50) is predicted.
\end{abstract}

%\keywords{}

\maketitle

\section{Introduction}

The long chain of calcium isotopes provides an ideal laboratory for both theoretical and experimental investigations of unstable nuclei. With two typical  doubly-magic isotopes $^{40}$Ca and $^{48}$Ca, the calcium chain is speculated to be up to  $^{70}$Ca, a possible third isotope of the double magicity. Current experiments have reached $^{60}$Ca \cite{PhysRevLett.121.022501}, but theoretical calculations are various \cite{erler2012limits,PhysRevC.80.044311,PhysRevLett.109.032502,PhysRevC.89.024319,PhysRevC.89.061301,PhysRevC.90.024312,PhysRevC.90.041302,PhysRevLett.118.032502}. Further refined calculations are still in demand. Besides the \textit{N} = 20 and 28 magic numbers, experiments have also given evidences of additional shell closures at $^{52}$Ca (\textit{N} = 32) \cite{wienholtz2013masses} and  $^{54}$Ca (\textit{N} = 34) \cite{steppenbeck2013evidence}.
It is still an open question whether the $N=40$ shell closure vanishes in the calcium chain. The spherical $N=40$ shell closure remains in the isotone $^{68}$Ni \cite{PhysRevLett.88.092501}, while it disappears in the isotones $^{64}$Cr and $^{66}$Fe with the onsets of deformation and collectivity \cite{PhysRevC.77.054306,PhysRevLett.102.012502,PhysRevC.81.061301,PhysRevLett.106.022502,PhysRevC.86.011305}.
 With advances in rare isotope beam facilities, more and more structure data will be obtained for calcium isotopes, which attracts  continuing interests of theory \cite{forssen2013living}.

 The  \textit{N} = 32 shell closure was observed in the early experiment \cite{PhysRevC.31.2226}, giving that $^{52}$Ca has a higher $2^+_1$ state by 1.5 MeV than in $^{50}$Ca. The precious mass measurements of  $^{53}$Ca and $^{54}$Ca  at CERN \cite{wienholtz2013masses}  show that the trend of two-neutron separation energies  supports the shell closure in $^{52}$Ca. The shell closure at $N=32$ has also been found in nearby titanium and chromium isotopes \cite{PhysRevLett.92.072502,PhysRevC.67.034309}. The spectroscopic experiment has reached  $^{54}$Ca, giving that the $2^+_1$ state in $^{54}$Ca is at 2.0 MeV  \cite{steppenbeck2013evidence}, slightly lower than in $^{52}$Ca, which provides an  experimental signature of  the shell closure in $^{54}$Ca. The precise  mass measurements of  $^{55{\text -}57}$Ca isotopes  provide additional experimental evidences for the understanding of the magic nature in $^{54}$Ca \cite{PhysRevLett.121.022506}.  To date, the mass measurements of calcium isotopes have been up to $^{57}$Ca,  while $^{54}$Ca is the heaviest calcium isotope for which spectroscopic data have been available. $^{60}$Ca is the neutron-richest calcium isotope obtained so far  in experiment \cite{PhysRevLett.121.022501}.  The experimental data provide valuable information to test theoretical calculations, and then lead to more reliable predictions for dripline nuclei and beyond.

The calcium region currently represents a frontier of theoretical calculations. With the phenomenological interactions, GXPF1A \cite{honma2005shell} and KB3G \cite{poves2001shell} for the \textit{pf} shell, large-scale shell-model calculations have been performed for calcium isotopes. The GXPF1A interaction results in a strong shell gap at $^{54}$Ca (\textit{N} = 34), while KB3G does not give the shell gap. Based on a realistic interaction of the CD-Bonn potential \cite{PhysRevC.63.024001}, the realistic shell model (RSM) with empirical single-particle (s.p.) energies \cite{PhysRevC.80.044311}  has been applied to  the spectra of calcium isotopes, predicting a weak shell closure at \textit{N} = 34. Further RSM calculations for the whole isotopic chain of calcium were done in Refs. \cite{PhysRevC.90.024312,PhysRevC.89.024319}. The complex coupled-cluster (CC) model \cite{PhysRevLett.109.032502,hagen2016emergent} with including the continuum effect has calculated up to $^{62}$Ca. In the complex CC calculations \cite{PhysRevLett.109.032502}, however, $^{60}$Ca is unbound, which is not consistent with the recent experiment \cite{PhysRevLett.121.022501}.
With a refined two- plus three-nucleon  $\Delta$NNLO interaction, the recent CC calculations extend the dripline beyond $^{60}$Ca \cite{Jiang:2020}.  Calcium isotopes have also been investigated by the Green's function (up to $^{52}$Ca) \cite{PhysRevC.89.061301} and in-medium similarity renormalization group (IM-SRG) (for even masses up to $^{62}$Ca) \cite{PhysRevC.90.041302} giving unbound  $^{56,58,60}$Ca. The nuclear density function theory (DFT) based on the Skyrme interaction predicts that the calcium two-neutron dripline should be at $^{70}$Ca \cite{erler2012limits,PhysRevLett.122.062502}.
An early work by the relativistic mean field  gave the dripline at $^{72}$Ca \cite{PhysRevC.65.041302}.
The theoretical calculations of  neutron-rich calcium isotopes are still a challenge, which needs  good understandings of the strong interaction, many-body correlation  and  coupling to scattering continuum.

In the present paper, we give the comprehensive calculations of neutron-rich calcium isotopes using the Gamow shell model GSM  with the CD-Bonn potential. The coupling to continuum is considered by using the complex-momentum (complex-$k$) Berggren space. In Sec.  \ref{two},  we describe the Berggren basis which treats bound, resonant and continuum states on equal footing. The effective Hamiltonian in the model space is derived from the realistic CD-Bonn interaction using the many-body perturbation theory (MBPT). The detailed calculations are given in Sec. \ref{three}. Binding energies, one- and two-neutron separation energies and excitation spectra are calculated and compared with existing data. The shell evolution in the calcium chain is discussed.

\section{Theoretical framework} \label{two}

 The Gamow resonance is a time-dependent problem, associated with particle emissions. To solve a time-dependent Schr\"{o}dinger equation  is extremely difficult, especially for many-body problems. Berggren \cite{berggren1968use} generalized the Schr\"{o}dinger equation to a complex-\textit{k}  plane in which the eigen energy is written as a complex number, $\widetilde e_i=e_i-i\gamma_i/2$, with $\gamma_i$ standing for the resonance width (measuring the half-life of particle emission). The Berggren method  provides an approach to solve a  time-dependent problem in a time-independent way.  In the complex-$k$ plane, the Berggren ensemble contains three types of states: bound, resonant and scattering continuum.

The shell model within the Berggren basis is called the Gamow shell model (GSM). With phenomenological interactions, the GSM has been successfully applied to the systems of two valence particles at first \cite{PhysRevLett.89.042501,PhysRevLett.89.042502} and more valence particles \cite{PhysRevC.67.054311,0954-3899-36-1-013101,PhysRevC.96.024308,PhysRevC.96.054316,PhysRevC.100.064303,PhysRevC.101.031301}.
 The GSM based on realistic nuclear forces has also been developed \cite{PhysRevC.73.064307,PhysRevC.80.051301,PhysRevC.88.044318,SUN2017227,PhysRevC.100.054313}. To calculate heavy nuclei, an inner core is usually taken in shell-model calculations. The harmonic oscillator (HO) basis is often adopted  in the conventional shell model, while the GSM usually uses the Woods-Saxon (WS) potential to create the Berggren basis \cite{PhysRevLett.89.042501,PhysRevLett.89.042502,PhysRevC.67.054311,0954-3899-36-1-013101,PhysRevC.96.024308,PhysRevC.96.054316,PhysRevC.100.064303,PhysRevC.101.031301,PhysRevC.73.064307,PhysRevC.80.051301,PhysRevC.88.044318,SUN2017227,PhysRevC.100.054313}.

For calcium isotopes, the magic $Z$ = 20 protons are well bound and hence can be treated in the HO basis. For neutrons, we use the spherical WS potential $V(r)=V_0/[1+e^{(r-R)/a}]$ with a spin-orbit coupling $V_{ls}(r)= - \chi\frac{1}{r}\frac{dV}{dr}\bf{l}\cdot\bf{s}$, to create the neutron Berggren basis, where $V_0=-V_{00}[1-\kappa (N-Z)/(N+Z)]$ and $R=r_0 A^{1/3}$. We fix $r_0=1.15$ fm and $a=0.63$ fm which were usually used in previous WS-type calculations (e.g., \cite{XU2000119}). In the GSM calculations of calcium isotopes, we choose the doubly magic $^{48}$Ca as the inner core, but for isotopes heavier than $^{60}$Ca the closed-shell $^{54}$Ca is taken as the core to reduce the model dimension and computational cost.
If we kept $^{48}$Ca as the core for the neutron-richest calcium isotopes, the model dimension  would be beyond the power of current computers when continuum channels are included.
The parameters $V_{00}$, $\kappa$ and $\chi$ are chosen such that the  neutron orbits $1p_{3/2}$, $1p_{1/2}$, and  $0f_{5/2}$ reproduce  experimental s.p. energies  obtained in $^{49}$Ca \cite{PhysRevC.93.044327} and  experimental one-neutron separation energy in $^{55}$Ca \cite{PhysRevLett.121.022506}.
Table \ref{sp} lists the WS single particle (s.p.) energies for valence neutrons in the shell-model space, obtained with $V_{00}$ = 62.8 MeV, $\kappa$ = 0.738 and $\chi$ = $0.593$. We see that the experimental neutron s.p. energies are well reproduced. The neutron $ 1p_{3/2}$, $1p_{1/2}$ and $0f_{5/2}$  orbits are bound, while $1d_{5/2}$ and $0g_{9/2}$ are resonant. The obtained one-neutron separation energy in $^{55}$Ca is 1.40 MeV which agrees with the experimental datum of $1.56(0.16)$ MeV \cite{PhysRevLett.121.022506}.

\begin{table}[h]
\centering
\caption{The WS neutron s.p. energies (in MeV) calculated with $^{48}$Ca or $^{54}$Ca, compared with experimental s.p. energies extracted from $^{49}$Ca \cite{PhysRevC.93.044327}.} % \label{tab:1} \\
\setlength{\tabcolsep}{0.2mm}{
\begin{tabular}{cccc}
\hline
\hline
 Neutron  & WS &Expt & WS  \\
 s.p. orbits & ($^{48}$Ca) &  & ($^{54}$Ca)\\
\hline

$1p_{3/2}$ & $-5.26$ & $-5.14$ & $-$  \\
$1p_{1/2}$ & $-3.05$ & $-3.11$ & $-$\\
$0f_{5/2}$ & $-1.15$ & $-1.15$ & $-1.40$ \\
$1d_{5/2}$ & $2.00-i0.86$ &  & $2.01-i0.93$\\
$0g_{9/2}$ & $2.37-i0.01$ &  & $2.35-i0.01$ \\

\hline
\hline
\end{tabular}}\label{sp}
\end{table}

The completeness of the Berggren ensemble requires to include  non-resonant  continuum channels described by contours ($L^+$) in the complex-$k$ plane \cite{PhysRevLett.89.042502,PhysRevLett.89.042501,0954-3899-36-1-013101,SUN2017227}. For a channel with narrow resonant state(s), the continuum contour $L^+$ is chosen to contain the narrow resonant state(s) \cite{SUN2017227}. For a continuum channel without narrow resonant state, the contour $L^+$ is chosen to be a segment lying on the real-momentum $x$-axis (starting from the origin of the coordinates) \cite{SUN2017227}.
 In numerical calculations, the continuum contour $L^+$ is discrete using the Gauss-Legendre quadrature method \cite{0954-3899-36-1-013101,PhysRevLett.89.042502,SUN2017227}. For a continuum channel without narrow resonant state, we set ten discretization points on the contour $L^+$, which has been well tested to be sufficient to get convergence. In our previous publication for the {\it sd}-shell nuclei \cite{SUN2017227}, we took eight discretization points that can give well converged results. The $0g_{9/2}$ orbit has a small imaginary part of the eigen energy (only 10 keV, almost bound). This state is very close to the real-momentum $x$ axis in the Berggren complex-momentum plane, thus a contour close to the $x$ axis is chosen for the $g_{9/2}$ channel with 18 discretization points.  The $1d_{5/2}$ orbit has a relatively large imaginary part of the energy. The $d_{5/2}$ continuum contour $L^+$ needs to include the $1d_{5/2}$ resonant state. For the $d_{5/2}$ contour, we set 44 discretization points. Note that a channel containing a resonant state which has a significant imaginary energy needs more discretization points to reach the  convergence of the numerical calculation. We have tested that such discretizations above provide well converged calculations for the mass region investigated. The detail about the Berggren continuum contour and discretizing can be found in the previous paper \cite{SUN2017227} in which less discretization points were taken for the {\it sd}-shell nuclei. In the present work, we focus on  neutron-rich calcium isotopes heavier than $^{48}$Ca. Neutrons are treated in the Berggren complex-$k$ basis. The active model space for the GSM calculations is the neutron {$\{$}$1p_{3/2}$, $1p_{1/2}$, $0f_{5/2}$, $0g_{9/2}$-resonant+continuum,  $1d_{5/2}$-resonant+continuum{$\}$} with $^{48}$Ca as core, while it is the neutron {$\{$}$0f_{5/2}$, $0g_{9/2}$-resonant+continuum,  $1d_{5/2}$-resonant+continuum{$\}$} with $^{54}$Ca as core. Effects from other partial waves (including the core polarization)  are included via many-body perturbation by the so-called  nondegenerate $\hat Q$-box folded diagrams \cite{SUN2017227}.

%  For bound states, their scattering continuum states can be chosen along the real momentum axis. On the contrary, for the resonance states, The contour $L^+$ has to be chosen to contain all the discrete narrow resonant states in the domain between $L^+$ and real $k$ axis.

 The intrinsic Hamiltonian of an $A$-body system reads
%\sum_{i<j}^{A}{V}_{\bf NN}^{(i,j)}-\frac{{p_i}\cdot{p_j}{2Am}}
   \begin{equation}\label{1}
    H=\sum_{i=1}^{A}\frac{{{\bm p_i}}^2}{2m}+\sum_{i<j}^{A}{V}_{\rm {NN}}^{(ij)}-\frac{\bm P^2}{2Am},
  \end{equation}
where $\bm{p}_i$ is the nucleon momentum in the laboratory coordinate,  and  $\bm P = \sum_{i=1}^{A} \bm {p}_i $ is the center-of-mass (c.m.) momentum of the system.  $V_{\rm NN}^{(ij)}$ is the nucleon-nucleon (NN) interaction.  The Hamiltonian can be rewritten with a one-body term and a residual two-body interaction via the introduction of an auxiliary one-body potential $U$,

\begin{eqnarray}
   H && = \sum_{i=1}^{A}(\frac{\bm {p}^2_i}{2m}+U)+\sum_{i<j}^{A}({V}_{\rm NN}^{(ij)}-U-\frac{\bm p_i^2}{2Am}-\frac{{\bm p_i}\cdot{\bm p_j}}{Am}) \nonumber \\
            && = H_0+H_1,
 \end{eqnarray}
 where  $H_0=\sum_{i=1}^{A}(\frac{\bm {p}^2_i}{2m}+U) $ has a one-body form, and $H_1=\sum_{i<j}^{A}({V}_{\rm NN}^{(ij)}-U-\frac{\bm p_i^2}{2Am}-\frac{{\bm p_i}\cdot{\bm p_j}}{Am})$ is the residual two-body interaction with the correction from the c.m. motion.

In the present work, $U$ is taken as the WS potential with parameters described above,  and ${V}_{\rm NN}^{(ij)}$ uses the CD-Bonn potential \cite{PhysRevC.63.024001}. To speed up the convergence of many-body calculations, usually the bare force is softened  to remove the strong short-range repulsive core. We  use the $V_{{\rm low}{\text -}k}$ method \cite{BOGNER20031} to soften the CD-Bonn by integrating out high-momentum components above a certain cutoff $\Lambda$. A hard cutoff of $\Lambda = 2.6$  $\rm fm^{-1}$ has been taken in the present calculations. A large $\Lambda$ can reduce the effect of the induced three-nucleon force (3NF)  \cite{PhysRevC.68.034320,SUN2017227}.  The $V_{{\rm low}{\text -}k}$ NN interaction is defined in a relative  momentum space, while the shell model is performed in the laboratory coordinate (e.g., HO basis). Therefore, a transformation from the relative and c.m. coordinates to the laboratory basis is needed. This procedure can be done conveniently by using the Brody-Moshinsky brackets \cite{MOSHINSKY1959104}.

In the HO basis, the two-body completeness relation is
\begin{equation}
\sum_{\alpha\leq\beta}|\alpha\beta\rangle\langle\alpha\beta|=\mathbf{1},
\end{equation}
where $|\alpha\beta\rangle$ is a two-particle state of the HO basis. The two-body interaction in the HO basis is given by
\begin{equation}
V_{\rm HO}=\sum_{\alpha\leq\beta}^{N_{\rm shell}} \sum_{\gamma\leq\delta}^{N_{\rm shell}} |\alpha\beta\rangle\langle\alpha\beta|V_{{\rm low}{\text -}k}|\gamma\delta\rangle\langle\gamma\delta|,
\end{equation}
where $N_{\rm shell} = 2n+l+1$ with $n$ and $l$ being the node number and orbital angular momentum of the HO orbit, respectively. $N_{\text{shell}}$ indicates a truncation, i.e.,  how many HO shells are included in the calculation. The interaction elements  need to be further  converted to the Berggren basis for the GSM calculation. This can be done by computing overlaps between the Berggren and HO basis wave functions,
\begin{equation}\label{transfor}
\langle ab|V|cd\rangle = \sum_{\alpha\leq\beta}^{N_{\rm shell}}\sum_{\gamma\leq\delta}^{N_{\rm shell}} \langle ab|\alpha\beta\rangle\langle\alpha\beta|V_{{\rm low}{\text -}k}|\gamma\delta\rangle\langle\gamma\delta|cd\rangle,
\end{equation}
where $|ab\rangle$ ($|cd\rangle$) is a two-particle state of the Berggren basis. For identical particles (proton-proton or neutron-neutron) the overlaps are calculated by
\begin{equation}\label{eq:expand}
\langle ab|\alpha\beta\rangle=\frac{\langle a|\alpha\rangle \langle b|\beta\rangle-(-1)^{J-j_\alpha-j_\beta}\langle a|\beta\rangle\langle b|\alpha\rangle}{\sqrt{(1+\delta_{ab})(1+\delta_{\alpha\beta})}},
\end{equation}
where $J$ is the total angular momentum of the two-particle state, while $j$ is the angular momentum of a single-particle basis state. For the proton-neutron coupling, the overlap is given by
\begin{equation}
\langle ab|\alpha\beta\rangle=\langle a|\alpha\rangle\langle b|\beta\rangle.
\end{equation}
The overlaps between one-body basis wave functions are obtained by
\begin{equation}\label{eq:overlap}
\langle a|\alpha\rangle=\int dr r^2 u_a(r)R_\alpha (r)\delta_{l_al_\alpha}\delta_{j_aj_\alpha}\delta_{t_at_\alpha},
\end{equation}
where $u(r)$ and $R(r)$ are the radial parts of the Berggren and HO basis wave functions, respectively, with $l$, $j$ and $t$ being the orbital, total angular momentum and isospin quantum number, respectively.

The wave functions of resonant and continuum states spread widely in space and themselves are not square integrable. The transformation defined by Eq.(\ref{transfor}) has in fact utilized the short-range nature of the nuclear force. The Gaussian dying-out property of the HO wave functions with distance makes the overlaps integrable without divergence. For long-range operators, such as the kinetic energy, using Eq.(\ref{transfor}) is unreasonable in principle. We use the exterior complex scaling technique \cite{GYARMATI1971523} to treat the kinetic energy (and ${\bm p_i}{\bm p_j}$ terms) in the Berggren basis. $N_{\rm shell}$  should be large enough to get results converged. In the present work, we take $N_{\rm shell}=2n+l+1=24$ with the limit of $l\leq 5$. We have tested that such a HO truncation is sufficient to reach the convergences of the calculations.

The  interaction matrix elements obtained in the Berggren basis are complex and non-Hermitian. We employ the MBPT named the full $\hat{Q}$-box folded-diagram method \cite{KUO197165} to construct the realistic GSM effective interaction in the defined model space. The complex-$k$ Berggren basis states are non-degenerate, therefore a non-degenerate $\hat Q$-box folded-diagram perturbation named the extended Kuo-Krenciglowa (EKK) method  \cite{PhysRevC.89.044003}  has been used.  Using the MBPT, we first calculate  the  $\hat{Q}$-box  in the  complex-$k$ Berggren basis, as follows
 \begin{eqnarray}
   \widehat{Q}(E) && = PH_1P+PH_1Q\frac{1}{E-QHQ}QH_1P \nonumber \\
            && = PH_1P+PH_1Q\frac{1}{E-QH_0Q}QH_1P+...,
 \end{eqnarray}
where $E$ is the starting energy. $P$ and $Q$ stand for the active model space and excluded space, respectively, with $P + Q = \textbf{1}$. The $\hat{Q}$-box is composed of irreducible valence-linked diagrams  \cite{HJORTHJENSEN1995125,CORAGGIO20122125}. In the present calculation,  $\hat{Q}$-box diagrams are calculated up to the  second order. The derivatives of the $\hat{Q}$-box  are defined as
 \begin{eqnarray} \label{5}
 \widehat{Q}_k (E) && = \frac{1}{k!}\frac{d^k\widehat{Q}(E)}{dE^k} \nonumber \\
            && = (-1)^kPH_1Q\frac{1}{(E-QHQ)^{k+1}}QH_1P,
 \end{eqnarray}
where $k$ presents the $k$-th derivative.

  The effective Hamiltonian $H_{\rm eff}$ can be constructed via \cite{TAKAYANAGI201161}
\begin{equation} \label{heff}
  \widetilde{H}_{\textrm{eff}}=\widetilde{H}_{\textrm{BH}}(E)+\sum_{k=1}^{\infty}\widehat{Q}_k(E){\widetilde{H}_{\textrm{eff}}},
\end{equation}
where $\widetilde{H}_{\rm eff}$ stands for $\widetilde{H}_{\textrm{eff}}=H_{\textrm{eff}}-E$, while $\widetilde{H}_{\textrm{BH}}(E)=H_{\textrm{BH}}(E)-E$ is the Bloch-Horowitz Hamiltonian shifted by an energy $E$, with
 \begin{eqnarray}
 H_{\textrm{BH}}(E) && = PH_0P+\widehat{Q}(E) \nonumber \\
            && = PH_0P+PH_1P+PH_1Q\frac{1}{E-QHQ}QH_1P.
 \end{eqnarray}
 The $\widetilde{H}_{\rm eff}$ is obtained by  iterating Eq.(\ref{heff}), which is equivalently to calculate the folded diagrams with including high-order contributions by summing up the  subsets of diagrams to finite order. The  effective Hamiltonian is given by $H_{\rm eff} = \widetilde{H}_{\rm eff}+ E$, and the effective interaction is obtained by $V_{\rm eff} = H_{\rm eff}-PH_0P$.

We choose $^{48}$Ca as the inner core and $\{1p_{3/2}, 1p_{1/2}, 0f_{5/2}$, $0g_{9/2}$-${\rm resonance+continuum}$, $1d_{5/2}$-${\rm resonance+continuum}\}$ as the model space for valence neutrons outside the $^{48}$Ca core.  For isotopes heavier than $^{60}$Ca, we take the closed-shell $^{54}$Ca as the inner core and the neutron $\{0f_{5/2}, 0g_{9/2}$-${\rm resonance+continuum}$, $1d_{5/2}$-${\rm resonance+continuum}\}$ as the model space to reduce the model dimension and computational task.
The non-Hermitian GSM Hamiltonian is diagonalized in the  model space by using the Lanczos method in the $m$ scheme. Due to the fact that the scattering continuum states are included in the shell-model space, the model dimension increases dramatically with increasing the number of particles in the continuum states \cite{0954-3899-36-1-013101}. Similar to our previous calculations \cite{SUN2017227},  we allow at most two particles in the continuum. It has been tested that such  truncation  can give a good convergence of the calculation.

\section{Calculations and discussions}\label{three}

\begin{figure}[!htb]
\includegraphics[width=0.8\columnwidth]{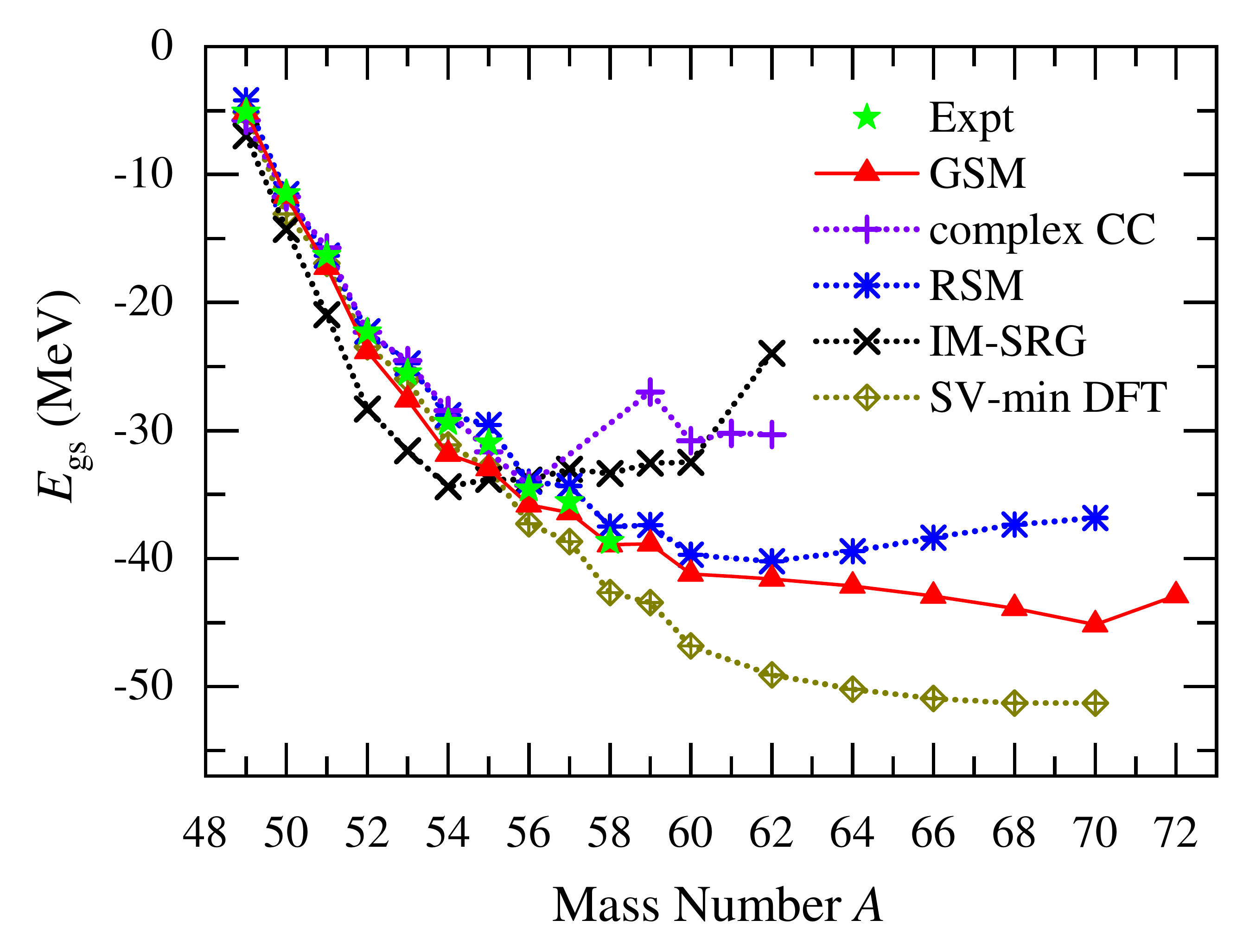}
\caption{Calculated ground-state energies with respect to $^{48}$Ca, compared with experimental data and other theoretical calculations: the complex CC with N$^3$LO(NN)+3NF$_{\text {eff}}$ \cite{PhysRevLett.109.032502}, RSM with N$^3$LO(NN)+NNLO(3NF) \cite{PhysRevC.90.024312}, IM-SRG with N$^3$LO(NN) + NNLO(3NF) \cite{PhysRevLett.118.032502} and SV-min DFT \cite{PhysRevC.79.034310}. The data for $^{48\text{-}54}$Ca have been collected in AME2016 \cite{wang2017ame2016}, while the data for $^{55{\text -}57}$Ca are taken from the recent experiment \cite{PhysRevLett.121.022506}. The $^{58}$Ca datum takes the evaluation given in AME2016 \cite{wang2017ame2016}. The CD-Bonn interaction is renormalized by $V_{{\rm low}{\text -}k}$ with $\Lambda=2.6$ fm$^{-1}$.}\label{Binding}
\end{figure}

Experiments to date have produced neutron-rich calcium isotopes up to $^{60}$Ca \cite{PhysRevLett.121.022501}, and mass measurements up to $^{57}$Ca \cite{PhysRevLett.121.022506}. We have made detailed calculations for isotopes up to $^{72}$Ca, using the GSM with the core and the corresponding model spaces described above. Figure \ref{Binding} shows the calculated ground-state energies, compared with experimental data \cite{PhysRevLett.121.022506,wang2017ame2016} and other theoretical calculations \cite{PhysRevLett.109.032502,PhysRevC.90.024312,PhysRevLett.118.032502,PhysRevC.79.034310}. There have existed several theoretical investigations within mean field (e.g., in \cite{PhysRevC.79.034310,PhysRevLett.122.062502}) and {\it ab initio}  (e.g., \cite{PhysRevLett.109.032502,PhysRevC.90.024312,PhysRevC.89.024319,PhysRevC.90.041302,PhysRevC.89.061301,PhysRevLett.118.032502}) models. The recent calculation based on the Skyrme-type DFT with the Bayesian statistical correction predicts that the neutron dripline would be at $^{70}$Ca \cite{PhysRevLett.122.062502}. The complex CC with chiral two-nucleon (NN) and density-dependent 3N forces  has calculated isotopes up to $^{62}$Ca \cite{PhysRevLett.109.032502}. The shell model with the Hamiltonian derived by MBPT based chiral NN and normal-ordered 3N forces has investigated the whole chain, giving slight decreases in binding energies beyond $^{62}$Ca \cite{PhysRevC.90.024312}. The IM-SRG with a chiral interaction has computed even-mass isotopes up to $^{62}$Ca \cite{PhysRevLett.118.032502}. We see in Fig. \ref{Binding} that the calculations lead to overall agreements in energies with existing data. The maximum discrepancy between the present calculation  and experimental energy is about 2.5 MeV happening in $^{54}$Ca.

\begin{figure}[!htb]
\includegraphics[width=0.8\columnwidth]{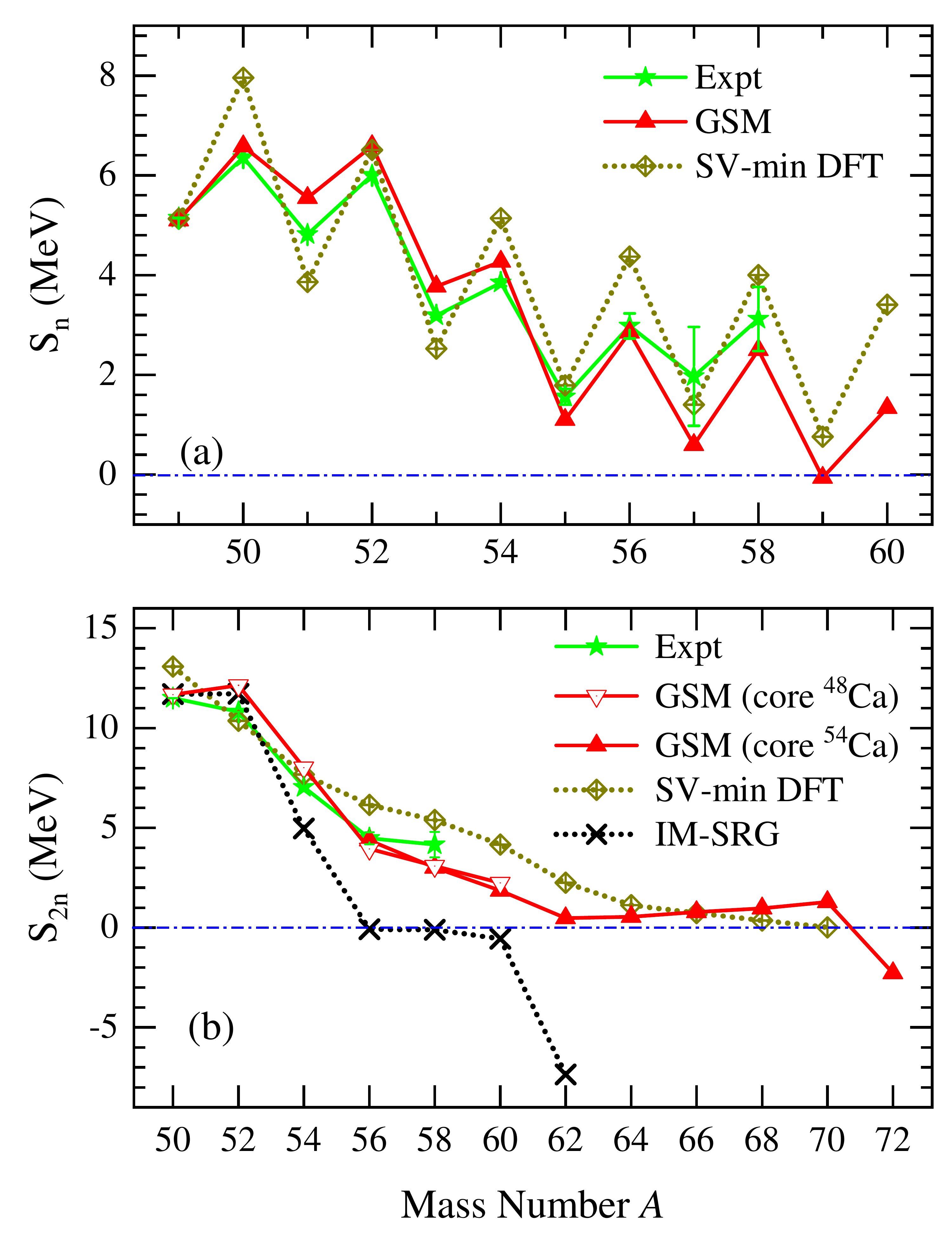}
\caption{Calculated one- (a) and two-neutron (b) separation energies, compared with data \cite{PhysRevLett.121.022506,wang2017ame2016} and calculations by SV-min DFT \cite{PhysRevC.79.034310} and multi-reference IM-SRG (only $S_{2n}$ calculated) \cite{PhysRevC.90.041302}.
The $S_n$ calculations stop at $^{60}$Ca because odd isotopes heavier than $^{60}$Ca become unbound in calculations.}\label{fig:2}
\end{figure}

Figure \ref{fig:2} displays one- and two-neutron  separation energies, compared with data \cite{PhysRevLett.121.022506,wang2017ame2016}, DFT \cite{PhysRevC.79.034310} and IM-SRG \cite{PhysRevC.90.041302} calculations.
The calculated one-neutron separation energies show that $^{57}$Ca is the heaviest odd-mass calcium isotope which is bound against neutron emission. This is consistent with the MBPT calculations \cite{PhysRevC.90.024312}.
$^{59}$Ca is weakly unbound with a small one-neutron separation energy of $-326$ keV in the present calculation. The experiment \cite{PhysRevLett.121.022501} observed a bound $^{59}$Ca. Theoretical predictions are various. The DFT calculation with the Bayesian statistical correction predicts that $^{59}$Ca is bound and $^{61}$Ca has a $\sim 50\%$ probability being bound \cite{PhysRevLett.122.062502}, while the relativistic mean-field calculation gives that the heaviest bound odd isotope is $^{59}$Ca \cite{PhysRevC.72.044318}. The IM-SRG calculations \cite{PhysRevLett.118.032502} show that the heaviest bound odd isotope would be in $^{53{\text -}59}$Ca.

Figure \ref{fig:2}(b) gives two-neutron separation energies in even calcium isotopes. To see whether the different choices of the shell-model core give consistent results, we have performed two kinds of calculations with the $^{48}$Ca or $^{54}$Ca core for $^{56,58,60}$Ca. We see in Fig. \ref{fig:2}(b) that the resulted two-neutron separation energies are well similar. The calculated two-neutron separation energies show an overall agreement with experimental data and other theoretical calculations, e.g., by DFT \cite{PhysRevC.79.034310}  and IM-SRG \cite{PhysRevC.90.041302}. The large two-neutron separation energies at $N$ = 32 and 34 indicate the subshell closures which have been suggested in experiments \cite{PhysRevC.31.2226,PhysRevC.74.021302,wienholtz2013masses,steppenbeck2013evidence,PhysRevLett.121.022506} and theories \cite{PhysRevC.80.044311,PhysRevLett.109.032502,PhysRevC.90.024312,PhysRevC.89.024319,PhysRevC.90.041302}. From the calculated two-neutron separation energies, we predict that the two-neutron dripline of the calcium chain should locate at $^{70}$Ca. This agrees with the recent mean-field calculation \cite{PhysRevLett.122.062502}.

\begin{figure}[!htb]
\includegraphics[width=0.8\columnwidth]{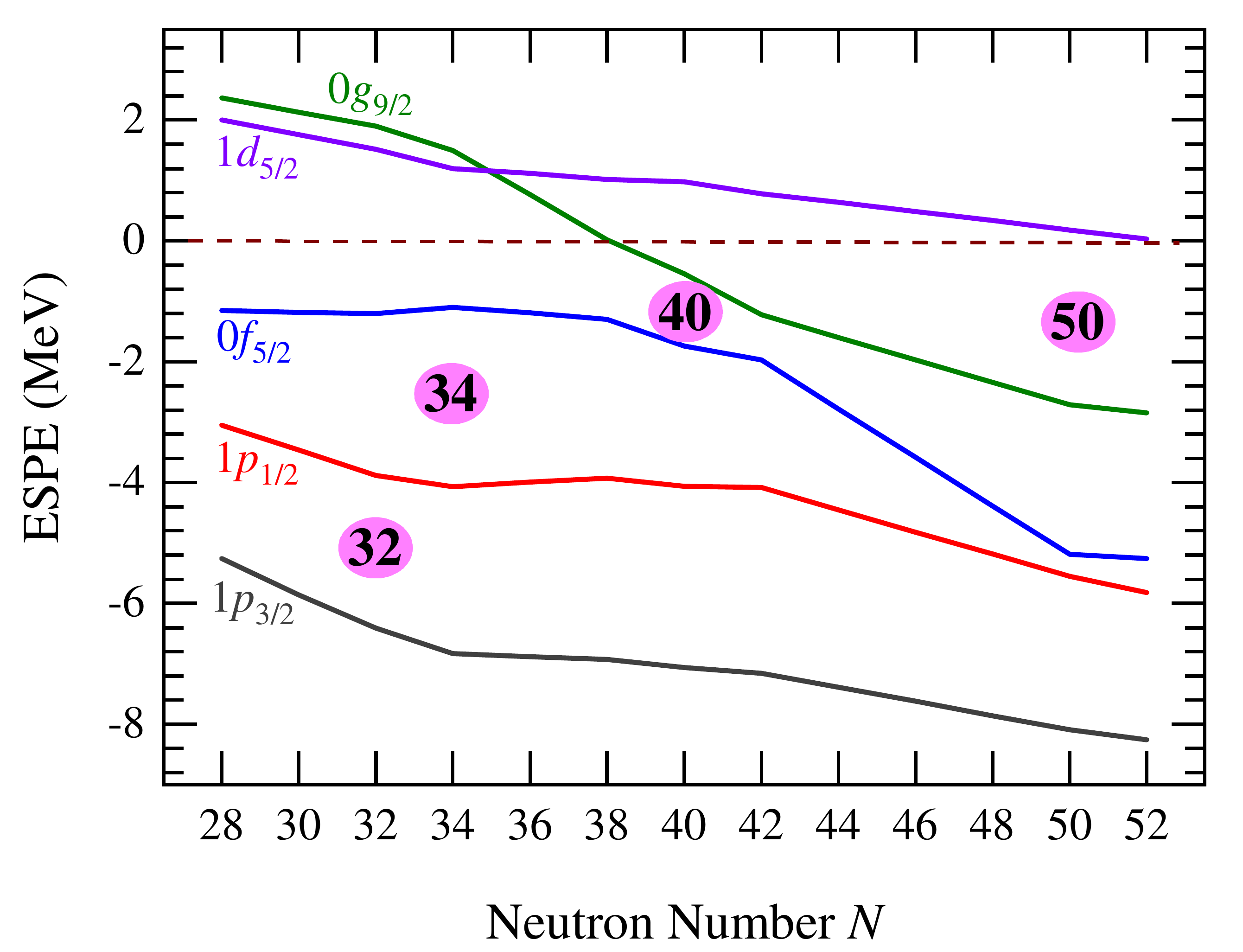}
\caption{Neutron effective single-particle energies (ESPE) with respect to the $^{48}$Ca core, as a function of neutron number. The $V_{\text{low-}k}$ $\Lambda = 2.6$ fm$^{-1}$ CD-Bonn interaction is used.}\label{fig:3}
\end{figure}

The  shell evolution in the calcium chain around the neutron numbers \textit{N} = 32, 34 and 40 is an interesting topic \cite{wienholtz2013masses,steppenbeck2013evidence,PhysRevLett.121.022506}. With the $V_{\text{low-}k}$ $\Lambda = 2.6$ fm$^{-1}$ CD-Bonn interaction, we have estimated the effective single-particle energy (ESPE) defined in Ref. \cite{PhysRevLett.87.082502}.  Figure \ref{fig:3} shows  the evolutions of the valence neutron  ESPEs with increasing the neutron number. We see that significant shell gaps exist between $1p_{3/2}$ and $1p_{1/2}$ and between $1p_{1/2}$ and $0f_{5/2}$, indicating shell closures at $N$ = 32 and 34, respectively. This is consistent with experimental observations \cite{PhysRevC.31.2226,PhysRevC.74.021302,wienholtz2013masses,steppenbeck2013evidence,PhysRevLett.121.022506} and theoretical calculations \cite{PhysRevC.80.044311,PhysRevLett.109.032502,PhysRevC.90.024312,PhysRevC.89.024319,PhysRevC.90.041302}.
The shell gap above the $0f_{5/2}$ orbit is reduced around $N = 40$, implying a weakening of the $N = 40$ shell closure in the calcium chain. In the isotone $^{68}$Ni the spherical $N = 40$ shell closure exists \cite{PhysRevLett.88.092501}, while the shell closure vanishes in the isotones $^{64}$Cr and $^{66}$Fe with the onset of deformation and collectivity \cite{PhysRevC.77.054306,PhysRevLett.106.022502,PhysRevC.86.011305,PhysRevC.81.061301,PhysRevLett.102.012502}.
The $N$ = 40 shell closure is eroded due to the intrusion of the $0g_{9/2}$ orbit, as illustrated in Fig. \ref{fig:3}.
The $0g_{9/2}$ orbit can drive the nucleus to be deformed. However, the onset of deformation depends upon how much $0g_{9/2}$ component appears in the state. The derived effective interaction gives a strong monopole attraction  between the $0f_{5/2}$ and $0g_{9/2}$ orbits, which results in the drop of the $0g_{9/2}$ orbit with increasing the neutron number. The occupation of the $0g_{9/2}$ orbit leads to the phenomenon of the so-called {\it island of inversion} predicted around $N$ = 40 in Cr and Fe isotopes \cite{PhysRevC.82.054301}. The $0g_{9/2}$ orbit becomes bound at $N \ge 40$, which can enhance the stability of heavy calcium isotopes.
The experimental discovery of the $^{60}$Ca ($N = 40$) \cite{PhysRevLett.121.022501} may be an indication of the enhanced stability. The ESPEs in Fig. \ref{fig:3} show a clear shell gap at $N = 50$, implying a shell closure there.

\begin{figure}[!htb]
\includegraphics[width=0.8\columnwidth]{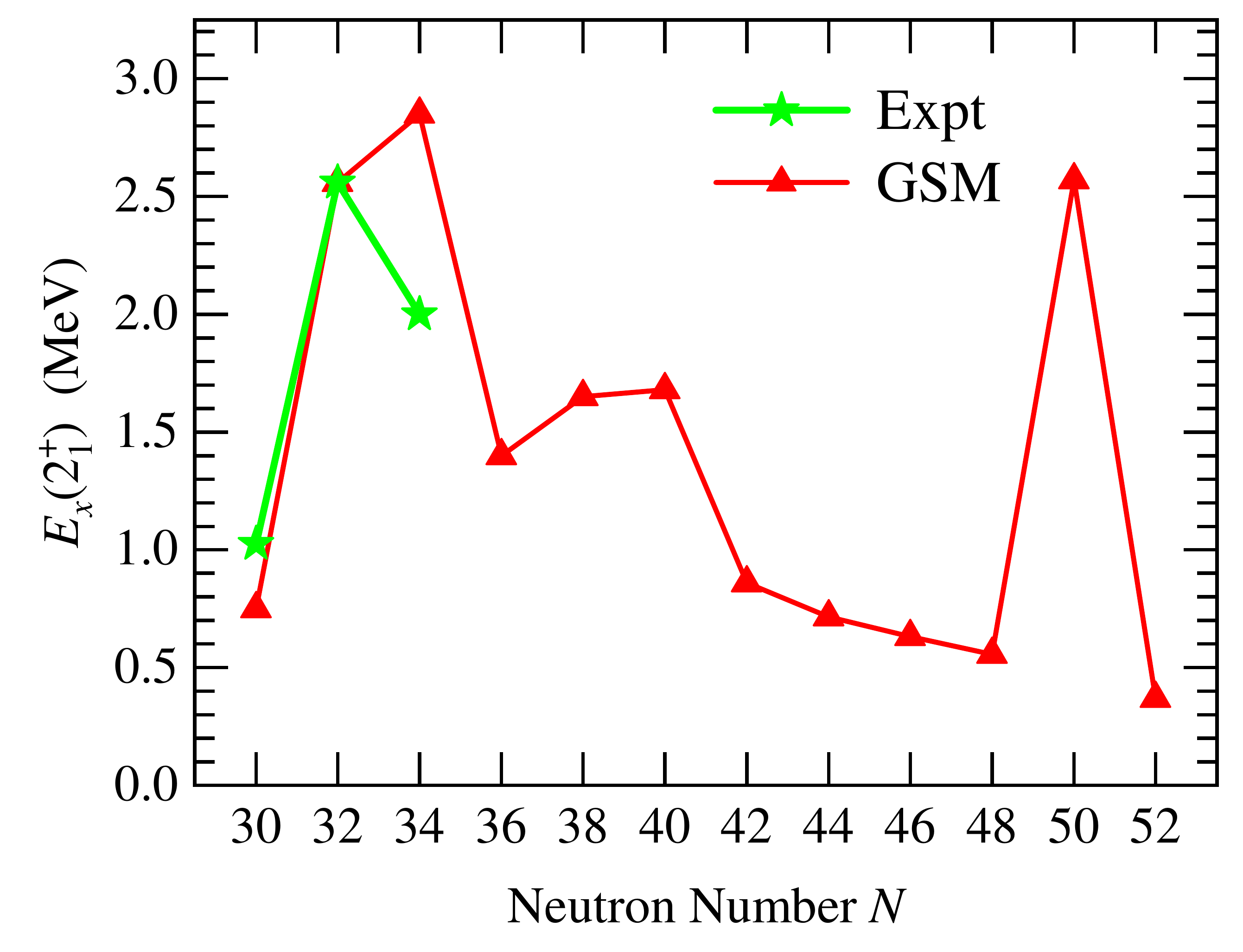}
\caption{Calculated excitation energies of the $2^+_1$ excited states for calcium isotopes, compared with data \cite{PhysRevC.74.014313,PhysRevC.76.021304,steppenbeck2013evidence}. }\label{fig:4}
\end{figure}

The excitation energy of the first $2^+$ excited state in even-even nuclei can be used to analyze the shell gap. With the present GSM based on the $V_{{\textrm {low-}}k}$ CD-Bonn interaction, we have calculated the excitation energies $E_x(2^+_1)$ for  neutron-rich calcium isotopes, shown in Fig. \ref{fig:4}. We see that at $N$ = 32 and 34 the obtained $2^+_1$ energies are significantly larger than in neighboring isotopes, which implies shell closures at $N$ = 32 and 34. This is consistent with the experimental  \cite{PhysRevC.31.2226,PhysRevC.74.021302,wienholtz2013masses,steppenbeck2013evidence,PhysRevLett.121.022506} and theoretical \cite{PhysRevC.80.044311,PhysRevLett.109.032502,PhysRevC.90.024312,PhysRevC.89.024319,PhysRevC.90.041302} conclusions. The calculated $2^+_1$ excitation energies around $N$ = 40 are lower than at $N$ = 32 and 34, which implicates a reduction of shell gap at $N$ = 40 in the calcium chain. The spherical $N$ = 40 shell closure was suggested experimentally in the isotone $^{68}$Ni with $E_x(2_1^+)=2033$ keV \cite{PhysRevLett.88.092501}, while the shell closure vanishes in the isotones, $^{66}$Fe with $E_x(2_1^+)=573$ keV \cite{PhysRevLett.82.1391} and $^{64}$Cr with $E_x(2_1^+)=420$ keV \cite{PhysRevC.81.051304}.
Experiments \cite{PhysRevC.77.054306,PhysRevLett.106.022502,PhysRevC.86.011305,PhysRevC.81.061301,PhysRevLett.102.012502} show more collectivity in $^{66}$Fe and $^{64}$Cr. The $2^+_1$ state in the lighter isotone $^{62}$Ti has not been detected in experiment. However, the experiment  observed a $2^+_1$ state in $^{60}$Ti ($N$ = 38) at an energy of 850 keV \cite{PhysRevLett.112.112503} which is about twice the excitation energy of the $2^+_1$ state in the isotone $^{62}$Cr. This would implicate possible higher $2^+_1$ excitation energies in $^{62}$Ti and $^{60}$Ca than in the  isotones $^{66}$Fe and $^{64}$Cr. Indeed, the shell-model calculations \cite{PhysRevC.82.054301} give higher $2^+_1$ energies in $^{60}$Ca and $^{62}$Ti than in the isotones $^{64}$Cr and $^{66}$Fe. Our calculation shown in Fig. \ref{fig:4} gives a $2^+_1$ state around 1.6 MeV for $^{60}$Ca. This result is consistent with the shell-model calculations in Refs. \cite{PhysRevC.89.024319,PhysRevC.82.054301}. Such a $2^+_1$ excitation energy is remarkably higher than in the isotones $^{64}$Cr and $^{66}$Fe which have the $2^+_1$ energies around  500 keV observed experimentally \cite{PhysRevC.81.051304,PhysRevLett.82.1391}. Though the $N = 40$ shell gap is reduced compared with the $N = 32$ and 34 gaps in the calcium chain, the sizable $2^+_1$ excitation energy in $^{60}$Ca would indicate an enhancement in the stabilities of $^{60}$Ca and heavier isotopes.

The GSM calculation gives that the dominant configurations of the $^{60}$Ca ground state are $\nu\{(1p_{3/2})^4(1p_{1/2})^2(0f_{5/2})^4(0g_{9/2})^2\}$ (50\%) and $\nu\{(1p_{3/2})^4(1p_{1/2})^2(0f_{5/2})^6\}$ (30\%), where the percentage indicates the proportion of the component.
We see that there is a $50\%$ probability of one pair of neutrons occupying the $0g_{9/2}$ intruder orbit. However, such an occupation in $0g_{9/2}$ should not be able to lead to a stable deformation in $^{60}$Ca which has a spherical proton magicity of $Z=20$. In Fig. \ref{fig:4}, we see a high $2^+_1$ excitation energy at $N$ = 50, which should indicate a shell closure there. The result is consistent with the shell-model calculation in Ref. \cite{PhysRevC.89.024319}. Combining the two-neutron separation energy given in Fig. \ref{fig:2}(b), we predict a doubly magic dripline nucleus of $^{70}$Ca for the calcium chain.

\begin{figure}[!htb]
\includegraphics[width=0.8\columnwidth]{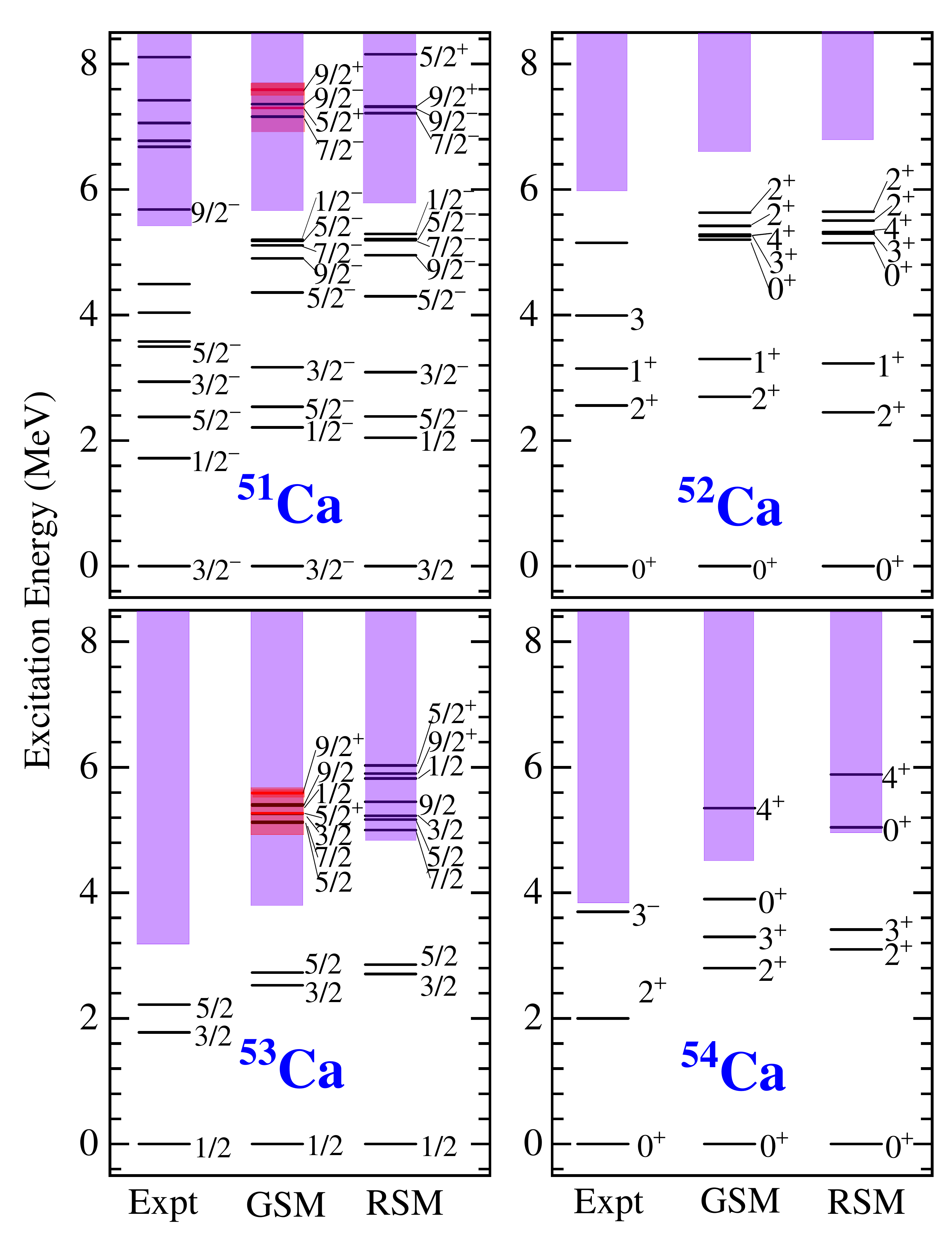}
\caption{Calculated excited states for $^{51{\text -}54}$Ca. The same $V_{\text{low-}k}$ $\Lambda = 2.6$ fm$^{-1}$ CD-Bonn interaction is used in GSM and RSM calculations. Data are from \cite{PhysRevC.74.014313,PhysRevC.76.021304,steppenbeck2013evidence}. Particle continua above emission thresholds are marked by purple shadowing, while resonant states are indicated by red shadowing.}\label{fig:5}
\end{figure}

Spectroscopic calculations can provide further information on nuclear structures. The experimental spectroscopy has reached  $^{54}$Ca \cite{steppenbeck2013evidence}. In Fig. \ref{fig:5}, we show the GSM calculations of excitation spectra for $^{51{\text-}54}$Ca, compared with experimental spectra available.  To see the effect from the continuum, we have also made conventional shell-model calculations within the HO basis, denoted by RSM as in Refs. \cite{PhysRevC.89.024319,PhysRevC.82.054301}.
The same $V_{\text{low-}k}$ $\Lambda = 2.6$ fm$^{-1}$ CD-Bonn potential is used.
The RSM space  for valence neutrons is $\lbrace 1p_{3/2}, 1p_{1/2},0f_{5/2},0g_{9/2},1d_{5/2} \rbrace$ which is the same as in the GSM calculation, except that the $g_{9/2}$ and $d_{5/2}$ continuum partial waves are not able to be included in the discrete HO basis. We see that low-lying excited states given by GSM and RSM are similar and agree with experimental data. This can be understood by the fact that the continuum effect is not significant in well-bound states. Resonances are seen in the GSM calculations around $E_{x}\sim 7.1$ and $\sim 5.1$ MeV (i.e., $\sim 1.6$ and $\sim 1.3$ MeV above one-neutron emission thresholds) for $^{51}$Ca and $^{53}$Ca, respectively. The resonant $5/2^+_1$ and $9/2^+_1$ excited states in the odd Ca isotopes reflect the resonant single-particle orbits of $1d_{5/2}$ and $0g_{9/2}$. The ordering of the $5/2^+_1$ and $9/2^+_1$ excited states is consistent with the order of the $1d_{5/2}$ and $0g_{9/2}$  orbits (as shown in Fig. \ref{fig:3}). Table \ref{resonance} gives the $5/2^+_1$ and $9/2^+_1$ resonant excited states predicted for the odd Ca isotopes. The $9/2^+$ state has a large $l = 4$ centrifugal barrier and hence a weak coupling to the continuum, giving a narrow resonance (or called quasi-bound state) \cite{PhysRevLett.109.032502}. By contrast, the $5/2^+$ state has a stronger coupling to continuum with a  lower $l = 2$ centrifugal barrier, resulting in a wide resonance.

\begin{table}[!htb]
\centering
\caption{Predicted $5/2^+_1$ and $9/2^+_1$ resonant states in odd calcium isotopes $^{51,53,55,57}$Ca by the GSM with the $V_{\text{low-}k}$ $\Lambda = 2.6$ fm$^{-1}$  CD-Bonn interaction. The excitation energy is defined by $\widetilde{E}=E-i\Gamma/2$, where the real part ($E$) of the energy gives the level position while the imaginary part defines the resonance width $\Gamma$. Both energy and width are in MeV.} % \label{tab:1} \\
\setlength{\tabcolsep}{1.2mm}{
%\begin{tabular}{l|S|r}
\begin{tabular}{ccccccccccc}
\hline
\hline
Nuclei & \multicolumn{2}{c}{$^{51}$Ca} & \multicolumn{2}{c}{$^{53}$Ca} & \multicolumn{2}{c}{$^{55}$Ca} &  \multicolumn{2}{c}{$^{57}$Ca} \\
         & $E$ & $\Gamma$ & $E$ & $\Gamma$ & $E$ & $\Gamma$ & $E$ & $\Gamma$ \\
%Nuclei & $E(^{51}\rm Ca)$&$\Gamma(^{51}\rm Ca)$&$E(^{53}\rm Ca)$&$\Gamma(^{53}\rm Ca)$&$E(^{55}\rm Ca)$&$\Gamma(^{55}\rm Ca)$&$E(^{57}\rm Ca)$&$\Gamma(^{57}\rm Ca)$ \\
%\hline

\hline
$5/2^+_1$ & 7.30 & 1.22 & 5.27 &  1.00 & 2.32& 0.58 & 1.62 &0.48 \\
$9/2_1^+$ & 7.59 & 0.04 & 5.59& 0.01  &2.60 & 0.00 & 1.18 & 0.00\\

\hline
\hline
\end{tabular}}\label{resonance}
\end{table}

\begin{figure}
\includegraphics[width=0.8\columnwidth]{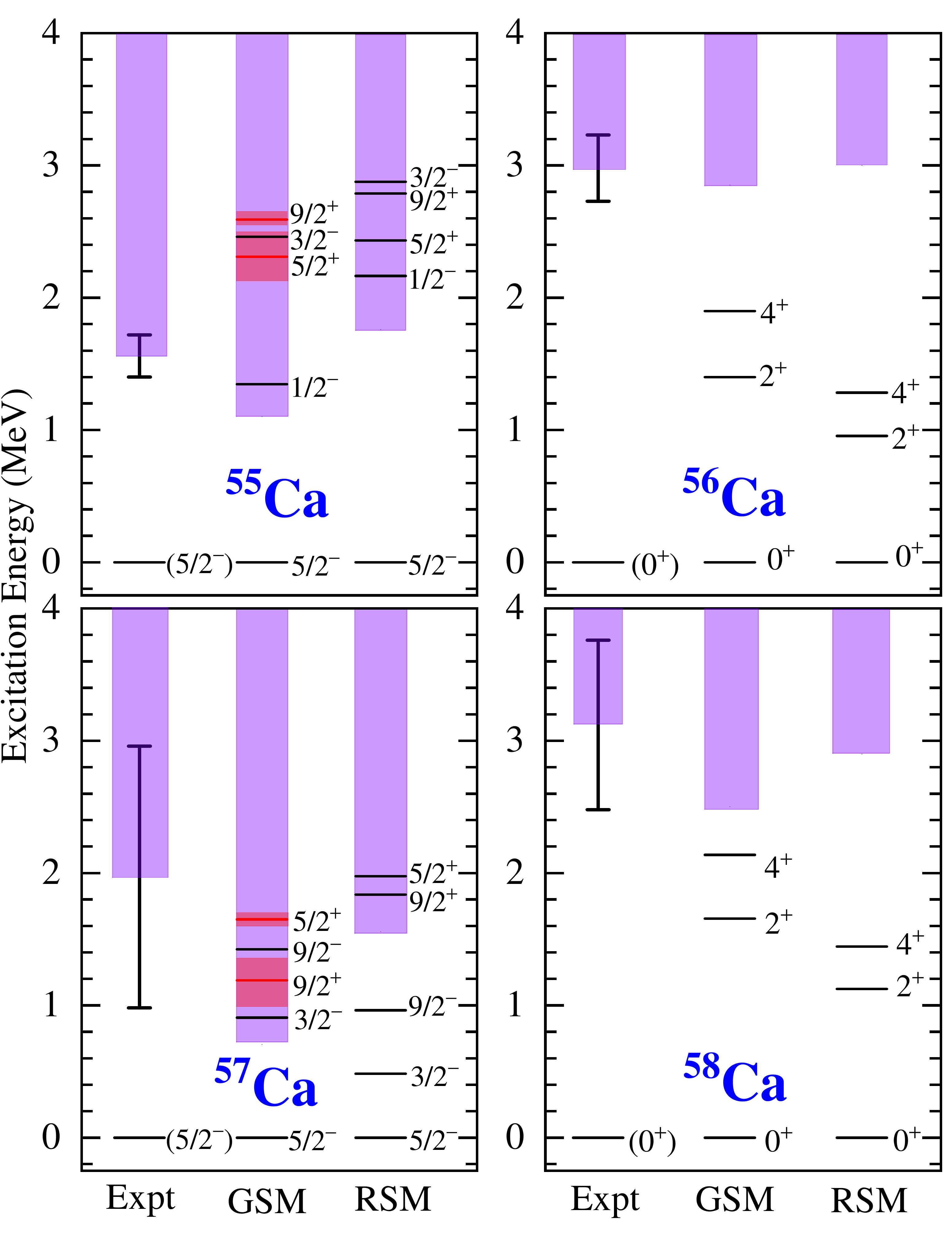}
\caption{Similar to Fig. \ref{fig:5}, but for $^{55\text -58}$Ca as the predictions of low-lying spectra. Note that the existing experimental data give large uncertainties in particle emission thresholds, indicated by error bars \cite{PhysRevLett.121.022506,wang2017ame2016}.}\label{fig:6}
\end{figure}

Figure \ref{fig:6} predicts low-lying excitation spectra in $^{55{\text -}58}$Ca, which should be useful for near future spectroscopic experiments. In heavy isotopes, the continuum effect becomes more significant.  As described above, the $^{48}$Ca core is used in the calculations of isotopes lighter than $^{61}$Ca. In $^{55}$Ca, the $5/2^-$ ground state is governed by the odd neutron occupying the $0f_{5/2}$ orbit above the $^{54}$Ca Fermi level. For the negative-parity $1/2^-$ and $3/2^-$ excited states, the dominant configuration is the  odd  $0f_{5/2}$ neutron  coupling to the first  $2^+$ excited state of $^{54}$Ca. The positive-parity $5/2^+$ and $9/2^+$ states have the odd neutron being excited to the $1d_{5/2}$ and $0g_{9/2}$ orbits, respectively. In $^{56}$Ca, the first $2^+$ and $4^+$ excited states are below the neutron emission threshold, with the dominant configuration of $(\nu 0f_{5/2})^2\otimes^{54}$Ca. In $^{57}$Ca, the $5/2^-$ ground state, $3/2^-$ and $9/2^-$ excited states are dominated by the $(\nu 0f_{5/2})^3\otimes^{54}$Ca configuration. The positive-parity $9/2^+$ and $5/2^+$ excited states have a character of single-particle excitation, with the odd neutron being excited to the $0g_{9/2}$ and $1d_{5/2}$, respectively. In $^{58}$Ca, the $0^+$ ground state and the first $2^+$, $4^+$ excited states contain two dominate configurations of $(\nu 0f_{5/2})^4\otimes^{54}$Ca and $(\nu 0f_{5/2})^2(\nu 0g_{9/2})^2\otimes^{54}$Ca with the intruder $0g_{9/2}$ orbit involved. We see that there exist low-lying resonant $5/2^+$ excited states in the odd isotopes $^{55,57}$Ca.

At the end, we test how sensitive the predictions are to the choice of interaction, by performing similar calculations but using different effective interactions: a softer CD-Bonn and the chiral N$^3$LO \cite{PhysRevC.68.041001,MACHLEIDT20111} with $\Lambda=2.3$ fm$^{-1}$ in $V_{\text{low-}k}$.
With the $^{54}$Ca core, the CD-Bonn GSM calculations with $\Lambda=2.3$ fm$^{-1}$ give that the ground states become slightly more bound by 0.4$-$2.9 MeV from $^{56}$Ca to $^{72}$Ca, compared with the calculations at $\Lambda=2.6$ fm$^{-1}$. It has been known that a soft interaction without 3NF invoked can lead to overbinding energies. The 3NF effect can be reduced by choosing a large $\Lambda$ cutoff in the $V_{\text{low-}k}$ procedure. The induced 3NF usually provides a repulsive effect on the binding energy, and the effect becomes larger as the number of valence particles increases. However, we find that the neutron separation energies which are the differences of binding energies do not change much from $\Lambda=2.6$ to 2.3 fm$^{-1}$. The conclusions remain unchanged with $\Lambda=2.3$ and 2.6 fm$^{-1}$, e.g., the heaviest bound odd isotope is $^{57}$Ca, and $^{70}$Ca remains the dripline nucleus. The calculations with $\Lambda=2.3$ and 2.6 fm$^{-1}$ give almost the same $2^+_1$ excitation energy in $^{70}$Ca (with a difference of only 0.1 MeV). The calculations using the chiral N$^3$LO softened with $\Lambda=2.3$ fm$^{-1}$ give the binding energies within 0.6 MeV of the $\Lambda=2.6$ fm$^{-1}$ CD-Bonn results. However, the heaviest bound odd isotope is $^{59}$Ca with a small one-neutron separation energy of only 0.08 MeV. This result seems to be consistent with the experiment \cite{PhysRevLett.121.022501} and the mean-field calculations \cite{PhysRevC.79.034310,PhysRevLett.122.062502}. $^{70}$Ca is still the neutron dripline nucleus, with a $2^+_1$ excitation energy of 2.3 MeV which is 0.4 MeV lower than that in the $\Lambda=2.6$ fm$^{-1}$ CD-Bonn calculation.

\section{Summary}
Using the Gamow shell model with the high-precision charge-dependent Bonn nucleon-nucleon interaction renormalized by the $V_{\text{low-}k}$  technique, we have performed comprehensive calculations for neutron-rich calcium isotopes up to beyond the neutron dripline. The coupling to continuum is included in the Gamow shell model by using the complex-momentum Berggren basis in which bound, resonant and continuum states are treated on equal footing. The Gamow shell model calculations can well describe the resonant properties of particle emission states in weakly-bound or unbound nuclei. Nuclear binding energies and neutron separation energies are calculated up to $^{72}$Ca, predicting that the heaviest odd bound isotope is $^{57}$Ca and the dripline locates at $^{70}$Ca. The calculations of the $2^+_1$ excitation and effective single-particle energies, combined with two-neutron separation energies, show the shell closures at $N = 32$, 34 and 50 and a shell weakening at $N = 40$. Calculated low-lying excitation spectra in $^{51{\text -}54}$Ca agree well with existing data. As predictions for near future spectroscopic experiments, we have calculated low excited states for $^{55{\text -}58}$Ca, providing useful information about the configurations of the low-lying states. Resonant excited states emerge in  odd isotopes $^{51,53,55,57}$Ca, which involve heavily the widely-resonant neutron $1d_{5/2}$ orbit. The continuum effect is seen  by the comparison between the Gamow and conventional shell-model calculations.

%     It has to be stressed that present  calculations employ only two-body  CD-Bonn interaction, not take into account 3NFs \cite{PhysRevC.98.044305}. Future works, we will include the 3NFs to analysis the  effects of 3NFs and improve the calculations.

\begin{acknowledgments}

  Valuable discussions with Z.H. Sun, N. Michel, M. Hjorth-Jensen, L. Coraggio, S.M. Wang, Y.Z. Ma and J.C. Pei are gratefully acknowledged. This work has been supported by the National Key R\&D Program of China under Grant No. 2018YFA0404401; the National Natural Science Foundation of China under Grants No. 11835001 and No. 11921006; the State Key Laboratory of Nuclear Physics and Technology, Peking University under Grant No. NPT2020ZZ01; and the CUSTIPEN (China-U.S. Theory Institute for Physics with Exotic Nuclei) funded by the U.S. Department of Energy, Office of Science under Grant No. de-sc0009971. We acknowledge the High-performance Computing Platform of Peking University for providing computational resources.

\end{acknowledgments}

\bibliography{Ca_revision}

\end{document}